\documentclass[letter,11pt]{article}
\pdfoutput=1
\usepackage{heppub} 
\allowdisplaybreaks
\usepackage{hyperref}
\usepackage{graphicx}
\usepackage{wasysym}
\usepackage{amsmath}
\usepackage{amssymb} 
\usepackage[normalem]{ulem}
\usepackage{xspace}
\usepackage{slashed}
\usepackage[utf8]{inputenc}
\usepackage[T1]{fontenc}
\usepackage{scalerel,stackengine}

\newcommand{\nn}{\nonumber}

\newcommand{\ml}{\mathcal}
\newcommand{\Tr}{\mathrm{Tr}}
\newcommand{\be}{\begin{equation}}
\newcommand{\ee}{\end{equation}}

\newcommand{\bs}{\boldsymbol}

\preprint{IQuS@UW-21-055}

\title{Simple Hamiltonian for Quantum Simulation of Strongly Coupled $2+1$D SU(2) Lattice Gauge Theory on a Honeycomb Lattice}

\author[1]{Berndt M\"uller}
\affiliation[1]{Department of Physics, Duke University, Durham, NC 27708, USA}

\author[2]{and Xiaojun Yao}
\affiliation[2]{InQubator for Quantum Simulation, Department of Physics, University of Washington, Seattle, WA 98195, USA}

\emailAdd{bmueller@duke.edu, xjyao@uw.edu}

\abstract{We find a simple spin Hamiltonian to describe physical states of $2+1$ dimensional SU(2) lattice gauge theory on a honeycomb lattice with a truncation of the electric field representation at $j_{\rm max}=\frac{1}{2}$. The simple spin Hamiltonian only contains local products of Pauli matrices, even though Gauss's law has been completely integrated out.
}

\begin{document}
\maketitle

\section{Introduction}
\label{sec:intro}
The idea of using one well understood quantum system to simulate another one that is less well understood has a long history~\cite{feynman1986quantum}. It became appealing to many research areas with the recent rapid development of quantum information technologies. In the area of nuclear and particle physics, quantum simulation has attracted significant and yet still growing research interests~\cite{PhysRevLett.110.125303,tagliacozzo2013simulation,Muschik:2016tws,Zohar:2016iic,Bender:2018rdp,Klco:2018kyo,Kaplan:2018vnj,rico2018so,Klco:2019evd,Davoudi:2020yln,Shaw:2020udc,celi2020emerging,Ciavarella:2021nmj,Ciavarella:2021lel,deJong:2021wsd,Kan:2021nyu,ARahman:2021ktn,Bauer:2021gek,Hayata:2021kcp,Farrell:2022wyt,Farrell:2022vyh,Yao:2022eqm,Ciavarella:2022qdx,Davoudi:2022uzo,Mueller:2022xbg,Grabowska:2022uos,Kane:2022ejm,Funcke:2022uwc,Angelides:2022pah,Davoudi:2022xmb,ARahman:2022tkr,Bauer:2022hpo,Florio:2023dke,Funcke:2023jbq,Zache:2023dko,Hayata:2023bgh,bauer2023quantum,Halimeh:2023wrx,PRXQuantum.3.040316,Zhang:2023hzr,PhysRevA.107.043312,Angelides:2023bme}, because of its potential to avoid the sign problem that obstructs traditional  numerical approaches to compute the real-time dynamics of gauge theories that form the cornerstone of the Standard Model.

Gauge theories are relativistic quantum field theories invariant under local gauge transformations. The local gauge invariance poses many challenges to efficiently and accurately simulate gauge theories on near-term quantum computers. In many Hamiltonian formulations of lattice gauge theories such as the Kogut-Susskind Hamiltonian~\cite{PhysRevD.11.395}, quantum link model~\cite{Chandrasekharan:1996ih,Brower:1997ha} and loop-string-hadron formulation~\cite{Raychowdhury:2018osk,Raychowdhury:2019iki,Kadam:2022ipf}, interactions are local, but not all local degrees of freedom correspond to physical states. Only states satisfying local gauge invariance (Gauss's law) are physical. As a result, noise in quantum hardware or errors introduced by quantum algorithms (such as the Trotterization errors) can lead to unphysical results in the simulation. Many generic error mitigation techniques such as the zero-noise CNOT extrapolation~\cite{PhysRevLett.120.210501,PhysRevX.8.031027,He:2020udd} are not sufficient to fully recover physical results due to the limited gate fidelity and systematic errors of algorithms~\cite{Klco:2019evd}. 

There have been many studies trying to address this problem, such as integrating out Gauss's law (see e.g., Refs.~\cite{Chakraborty:2020uhf,Honda:2021aum}), adding a gauge violation penalty term~\cite{PhysRevLett.107.275301,PhysRevLett.109.125302,PhysRevLett.110.055302,PhysRevX.3.041018,PhysRevLett.109.175302,PhysRevD.95.094507,Halimeh:2019svu,Halimeh:2021vzf}, averaging over different gauge choices from a dynamical drive and quantum control (the so called ``dynamical decoupling''~\cite{Kasper:2020owz}), using symmetry protection~\cite{Tran:2020azk} and post-selection~\cite{Nguyen:2021hyk}, and mapping local Gauge invariance into conservation laws on specific quantum hardware~\cite{Zohar:2012xf,Zohar:2013zla,Zohar:2015hwa}. However, many of these methods have limitations: Integrating out Gauss's law completely is hard and, when possible, it usually leads to non-local interactions that are less efficient to simulate on quantum hardware without all-to-all connections. Gauge violation penalty terms that are quadratic  contain two-body interactions that are not necessarily local, and approaches with one-body penalty terms~\cite{Halimeh:2020ecg} require the preparation of an initial physical state, which can itself require solving the gauge error problem. Using symmetry protection and post-selection does not necessarily reduce errors since symmetry-preserving and symmetry-violating errors can interfere destructively and thus removing only the symmetry-violating error can increase the error~\cite{Nguyen:2021hyk}. Mapping gauge invariant interactions onto processes on quantum hardware that are protected by conservation laws are constrained to specific types of interactions and certain quantum technology platforms. Therefore, it would be extremely useful if one could find a generic formulation that is local but does not suffer from the problem of the error-induced admixture of unphysical states for universal quantum computers. 

Here, we provide such an example for $2+1$ dimensional SU(2) gauge theory at strong coupling. By considering a honeycomb lattice and truncating the electric basis at $j_{\rm max}=\frac{1}{2}$, we are able to map bijectively the $2+1$D SU(2) lattice gauge theory onto a $2$D spin model with local interactions. In this procedure, the local gauge invariance, i.e., Gauss's law is fully accounted for and only the physical Hilbert space is included in the description, which means time evolution driven by the constructed Hamiltonian is robust against error-induced unphysical states. The simple Hamiltonian will also enable a numerical test of the eigenstate thermalization hypothesis (ETH) for part of the Hilbert space of the $2+1$D SU(2) theory, as a two-dimensional extension of the plaquette chain study~\cite{Yao:2023pht}. 

This paper is organized as follows: In Section~\ref{sec:H} we will present the Hamiltonian of $2+1$ dimensional SU(2) gauge theory on a honeycomb lattice and describe how  it can be mapped onto a $2$D spin model when the electric field Hilbert space is truncated to $j_{\rm max}=\frac{1}{2}$. In Section~\ref{sec:k-sector} we will construct momentum eigenstates in the case of periodic boundary condition and calculate matrix elements of the Hamiltonian and Wilson loop operators. A brief summary will be given in Section~\ref{sec:conclusions}.

\section{2+1D SU(2) Gauge Theory on Honeycomb Lattice}
\label{sec:H}

\subsection{Kogut-Susskind (KS) Hamiltonian}
The Hamiltonian density of $2+1$D SU(2) gauge theory in the continuum can be written as
\begin{align}
\ml{H} = \frac{1}{2g^2}E_i^aE_i^a + \frac{1}{4g^2}F_{ij}^aF_{ij}^a \,,
\end{align}
where $E_i^a$ denotes the electric field along the $i$-th spatial direction with the SU(2) adjoint index $a\in{1,2,3}$, $F_{ij}^a$ represents the non-Abelian magnetic field (field strength tensor), and $g$ is the coupling with mass dimension $[g]=0.5$. Compared with the standard notation in the continuum, we have absorbed a factor of $g$ into the definition of $F_{\mu\nu}$. Later, we will absorb another factor of $a/g^2$ into the definition of the electric field where $a$ is the lattice spacing.

Now we want to construct a lattice version of the Hamiltonian on a 2D plane. For the standard square lattice  each vertex connects four links. A physical state at the vertex then cannot be uniquely defined by the $j$ values on the four links, where $j$ denotes the label of an electric basis state, and a fifth $j$ value is required to define a unique vertex state. Alternatively, when each vertex connects only three links as in the case of a square plaquette chain, physical vertex states can be uniquely determined by the corresponding $j$ values. This motivates us to consider a honeycomb lattice in two spatial dimensions.

\begin{figure}
\centering
\includegraphics[width=0.5\textwidth]{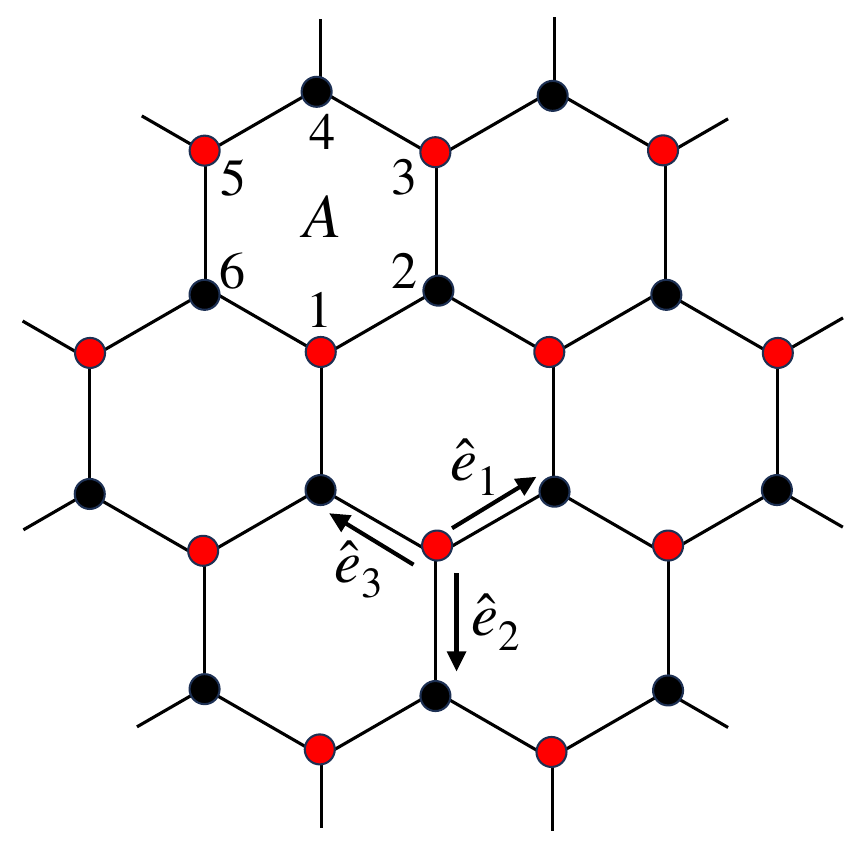}
~
~
\includegraphics[width=0.4\textwidth]{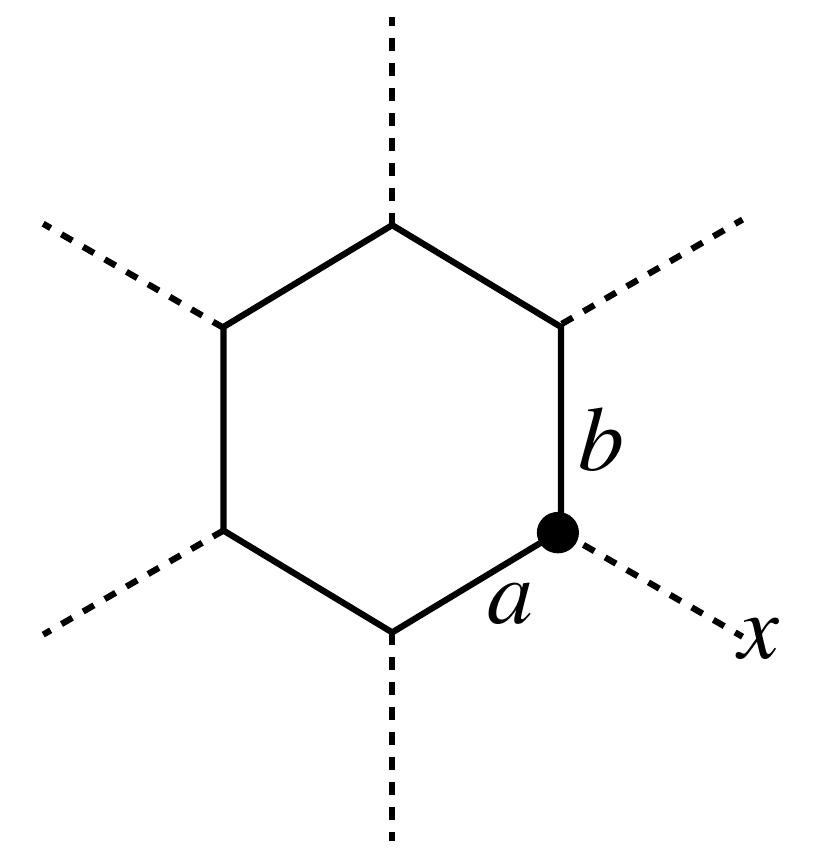}
\caption{Left panel: Honeycomb lattice in 2 spatial dimensions with three unit vectors defined pointing from a red dot to the three adjacent black dots. Right panel: Single plaquette shown by solid lines with six external links depicted by dashed lines. The black dot vertex is labeled as $(a,x,b)$.}
\label{fig:honeycomb}
\end{figure}

As shown in Fig.~\ref{fig:honeycomb}, the vertices defining a honeycomb lattice can be labeled as black or red such that no link connects two black or two red vertices. In order to formulate the discretized gauge field Hamiltonian on the honeycomb lattice we define three unit vectors pointing from a red site to an adjacent black site 
\begin{align}
\hat{e}_1 = \Big( \frac{\sqrt{3}}{2}, \frac{1}{2} \Big)\,,\qquad  \hat{e}_2 = (0, -1) \,, \qquad \hat{e}_3 = \Big( - \frac{\sqrt{3}}{2}, \frac{1}{2} \Big)\,, 
\end{align}
which satisfy $\sum_i \hat{e}_i=(0,0)$. We represent Wilson lines (link variables) that start at a red dot and point along these directions to be $U$. Those link variables that point in the opposite directions and end on a red dot are labeled as $U^\dagger$. For example, the link variable from vertex 1 to vertex 2 in plaquette $A$ can be written as
\begin{align}
U({\bs n}_1,\hat{e}_1) = \exp[ia \hat{e}_1\cdot {\bs A}({\bs n}_1) ]= \exp \bigg[ ia\Big( \frac{\sqrt{3}}{2}A_x({\bs n}_1) + \frac{1}{2}A_y({\bs n}_1) \Big) \bigg] \,,
\end{align}
where $a$ is the lattice spacing between connected black and red dots. The gauge field ${\bs A}\equiv {\bs A}^a T^a \equiv {\bs A}^a\sigma^a/2$ is a SU(2) matrix (here the superscript $a$ is a SU(2) adjoint index implicitly summed over and $\sigma^a$ denotes a Pauli matrix). The argument of the gauge field
${\bs n}_1$ denotes the position of vertex 1 in plaquette A. The link variable from vertex 2 to vertex 3 can be written as
\begin{align}
U^\dagger({\bs n}_3,\hat{e}_2) = \exp[-ia \hat{e}_2\cdot {\bs A}({\bs n}_3) ] = \exp[ ia A_y({\bs n}_3) ] \,,
\end{align}
where ${\bs n}_3$ denotes the position of vertex 3 in plaquette A.
Similarly the remaining link variables along a hexagonal plaquette are given by
\begin{align}
U({\bs n}_3,\hat{e}_3) 
& = \exp[ia \hat{e}_3\cdot {\bs A}({\bs n}_3) ] \nn
\\
U^\dagger({\bs n}_5,\hat{e}_1) 
& = \exp[-ia \hat{e}_1\cdot {\bs A}({\bs n}_5) ] \nn
\\
U({\bs n}_5,\hat{e}_2) 
& = \exp[ia \hat{e}_2\cdot {\bs A}({\bs n}_5) ] \nn
\\
U^\dagger({\bs n}_1,\hat{e}_3) 
& = \exp[ -ia \hat{e}_3\cdot {\bs A}({\bs n}_1) ] \,.
\end{align}

Expanding in powers of $a$, the plaquette operator at $A$ is given by
\begin{align}
\varhexagon_A & = \Tr[U^\dagger({\bs n}_1,\hat{e}_3)
U({\bs n}_5,\hat{e}_2)
U^\dagger({\bs n}_5,\hat{e}_1)
U({\bs n}_3,\hat{e}_3)
U^\dagger({\bs n}_3,\hat{e}_2)
U({\bs n}_1,\hat{e}_1)] \nn \\
& = \Tr \left[ 1 + ia^2 \frac{3\sqrt{3}}{2} F_{xy}({\bs n}_1) 
- \frac{1}{2}a^4 \frac{27}{4} F^2_{xy}({\bs n}_1) + \cdots \right] \,.
\end{align}
At lowest order in the lattice spacing $a$ we therefore obtain the result
\begin{align}
\Tr [F^2_{xy}({\bs n}_1)] = \frac{4}{27a^4} [4 - \varhexagon({\bs n}_1) - \varhexagon^\dagger({\bs n}_1)]\,.
\end{align}
Using the fact that the plaquette variable in SU(2) is real, $\varhexagon = \varhexagon^\dagger$, the magnetic energy part of the KS Hamiltonian can be written as
\begin{align}
H_{\rm mag} = \int {\rm d}^2x \frac{1}{4g^2}F^a_{ij}({\bs x})F^a_{ij}({\bs x})
= \frac{3\sqrt{3}}{2} a^2\sum_{\bs n} \frac{1}{g^2} \Tr[F^2_{xy}({\bs n})] 
= \frac{4\sqrt{3}}{9g^2a^2} \sum_{\bs n} \big(2-\varhexagon({\bs n})\big)\,,
\label{eq:Hmag}
\end{align}
where the sum ${\bs n}$ runs over the red dots on the honeycomb lattice and the factor $3\sqrt{3}/2$ comes from the area of the hexagon or the parallelogram formed by four nearest black or red dots. 

For the electric part of the Hamiltonian we decompose the two-component electric field into three parts
\begin{align}
E_1^a = \hat{e}_1\cdot {\bs E}^a\,,\qquad E_2^a = \hat{e}_2\cdot {\bs E}^a\,,\qquad E_3^a = \hat{e}_3\cdot {\bs E}^a\,,
\end{align}
which satisfy $({\bs E}^a)^2 = (E_x^a)^2+(E_y^a)^2 = (E_1^a)^2+(E_2^a)^2+(E_3^a)^2$. Thus, the electric part of the Hamiltonian is
\begin{align}
H_{\rm el} = \frac{g^2}{2} \frac{3\sqrt{3}}{2}\sum_{\bs n} \sum_{i=1}^{3} \sum_{a=1}^3 (E_i^a)^2(\bs n) \,,
\label{eq:Hel}
\end{align}
where again the sum over ${\bs n}$ denotes either the black or the red vertices of the honeycomb lattice and the factor $3\sqrt{3}/2$ again comes from the hexagon area.

The electric fields can be made to live either on black vertices or red vertices without changing the above form of $H_{\rm el}$. The commutation relations between $E$ and $U$ can be easily written out if we make the electric fields live on links rather than on black or red vertices. Then there will be two types of electric fields living on the same link: One induces gauge transformation of the link variable on the black end while the other induces gauge transformation on the red end. Either one can be used in $H_{\rm el}$. All in all, we can write
\begin{align}
\label{eq:EB}
[E^a_{Bi}({\bs n}+\hat{e}_i/2), U({\bs n},\hat{e}_j)] = -\delta_{ij} T^a U({\bs n},\hat{e}_j)  \,,
\end{align}
where $B$ denotes the electric field that is gauged on a black vertex. The argument of the electric field indicates it lives on a link. Similarly we have
\begin{align}
\label{eq:ER}
[E^a_{Ri}({\bs n}+\hat{e}_i/2), U({\bs n},\hat{e}_j)] = -\delta_{ij} U({\bs n},\hat{e}_j) T^a \,,
\end{align}
where $R$ denotes the electric field that is gauged on a red vertex. The ``black'' and ``red'' electric fields are the analogues of ``left'' and ``right'' electric fields on a square lattice. They are generators of local gauge transformation and have non-trivial commutation relations\footnote{The signs on the right hand sides of Eq.~\eqref{eq:[EE]} are consistent with those of Eqs.~\eqref{eq:EB} and~\eqref{eq:ER}, which are obtained from the commutation relation between an electric field and a Wilson line in the continuum. If we were working with a square lattice, it would be more convenient to redefine $E^a_{Ri}$ with $-E^a_{Ri}$ (On a square lattice, ``red'' corresponds to ``right''. In this way, the negative sign on the right hand sides of Eq.~\eqref{eq:ER} and the last line of Eq.~\eqref{eq:[EE]} would disappear. For a square lattice, this is more convenient since each vertex contains both ``left'' and ``right'' types of electric fields and redefining them such that they transform link variables in the same way simplifies the construction of singlets at vertices. On the other hand, each vertex on a honeycomb lattice only involves one type of electric fields, either ``black'' or ``red'' and thus the construction of singlets at each vertex is already simple.}
\begin{align}
\label{eq:[EE]}
[E^a_{Bi}({\bs n}+\hat{e}_i/2), E^b_{Bj}({\bs m}+\hat{e}_j/2)] &= i\varepsilon^{abc} \delta_{ij} \delta_{{\bs n}{\bs m}} E^c_{Bi}({\bs n}+\hat{e}_i/2) \nn\\
[E^a_{Ri}({\bs n}+\hat{e}_i/2), E^b_{Rj}({\bs m}+\hat{e}_j/2)] &= -i\varepsilon^{abc} \delta_{ij} \delta_{{\bs n}{\bs m}} E^c_{Ri}({\bs n}+\hat{e}_i/2) \,,
\end{align}
where $\varepsilon^{abc}$ is the Levi-Civita symbol and serves as the structure constant of the SU(2) group.

Altogether, the KS Hamiltonian on the honeycomb lattice reads
\begin{align}
H_{\rm KS} = \sum_{\bs n} \left( \frac{3\sqrt{3}g^2}{4}  \sum_{i=1}^{3} \sum_{a=1}^3 (E_i^a)^2(\bs n)
+ \frac{4\sqrt{3}}{9 g^2a^2}  \big(2-\varhexagon({\bs n})\big) \right) \, .
\label{eq:HKS}
\end{align}
The constant in the magnetic energy represents an overall shift of all energy eigenvalues and is often omitted in the literature. In addition to the Hamiltonian, physical states $|\psi_{\rm phy}\rangle $ satisfy Gauss's law 
\begin{align}
\sum_{i=1}^3 E_{i}^a |\psi_{\rm phy}\rangle = 0 \,,
\end{align}
for all $a$'s and all black and red dots.

The matrix elements of the KS Hamiltonian can be easily evaluated in the electric field representation, where the matrix elements of the electric energy operator are diagonal \cite{Byrnes:2005qx,Zohar:2014qma,Klco:2019evd,ARahman:2021ktn}. Denoting the SU(2) representation of the electric field by $j=0,{\frac{1}{2}},1,\ldots$ the contribution from each link is given by
\begin{align}
\langle J | \sum_{a=1}^3(E^a)^2| j\rangle = j(j+1)\delta_{Jj} \, .
\end{align}
The matrix elements of the plaquette operator can be evaluated using the techniques described in \cite{Hayata:2023puo}:
\begin{align}
\label{eq:6j}
\langle \{J\} | \varhexagon |\{j\} \rangle \equiv \langle \{J\} | \prod_{V=1}^{6} M_V |\{j\} \rangle 
= \prod_{V=1}^{6} (-1)^{j_a+J_b+j_x} \sqrt{(2J_a+1)(2j_b+1)}
\bigg\{ \begin{array}{ccc}  j_x & j_a & j_b \\ \frac{1}{2} & J_b & J_a   \end{array}  \bigg\} \, ,
\end{align}
where $M_V$ denotes the part of the honeycomb plaquette operator at the vertex $V =(a,x,b)$, where $a,b$ are the two internal plaquette links attached to the vertex, and $x$ denotes the external link at the vertex, as shown in Fig.~\ref{fig:honeycomb}. The symbol $\{j\}$ is a collection of the six initial $j$ values of the plaquette links (the $a,b$ type) while $\{J\}$ labels those of the final state, after the honeycomb plaquette operator has been applied. The $j_x$ value of the external link does not change. The curly bracket containing six $j$-values in two rows denotes the Wigner $6j$-symbol whose explicit expression can be found in Refs.~\cite{alma99116836810001452,alma99132772390001452}.

\subsection{Map to Spin Model with Truncation at $j_{\rm max}=\frac{1}{2}$}

We now discuss the truncation of the Hilbert space for the KS Hamiltonian to the electric field representations $j=0,\frac{1}{2}$. While this truncation would not yield meaningful results for physically interesting observables when the coupling constant is small, it allows us to search for ETH-type behavior for modestly large lattices. If the gauge theory constrained to this reduced Hilbert space exhibits quantum chaos, it appears plausible that this property is also present in the unconstrained Hilbert space. Furthermore, this truncation provides an example where Gauss's law is completely integrated out and yet the interactions are still local. In this sense, the model with this truncation provides a good starting point and benchmark for future quantum simulation studies of lattice gauge theories that are robust against unphysical states induced by errors.

\begin{figure}
\centering
\includegraphics[width=0.7\textwidth]{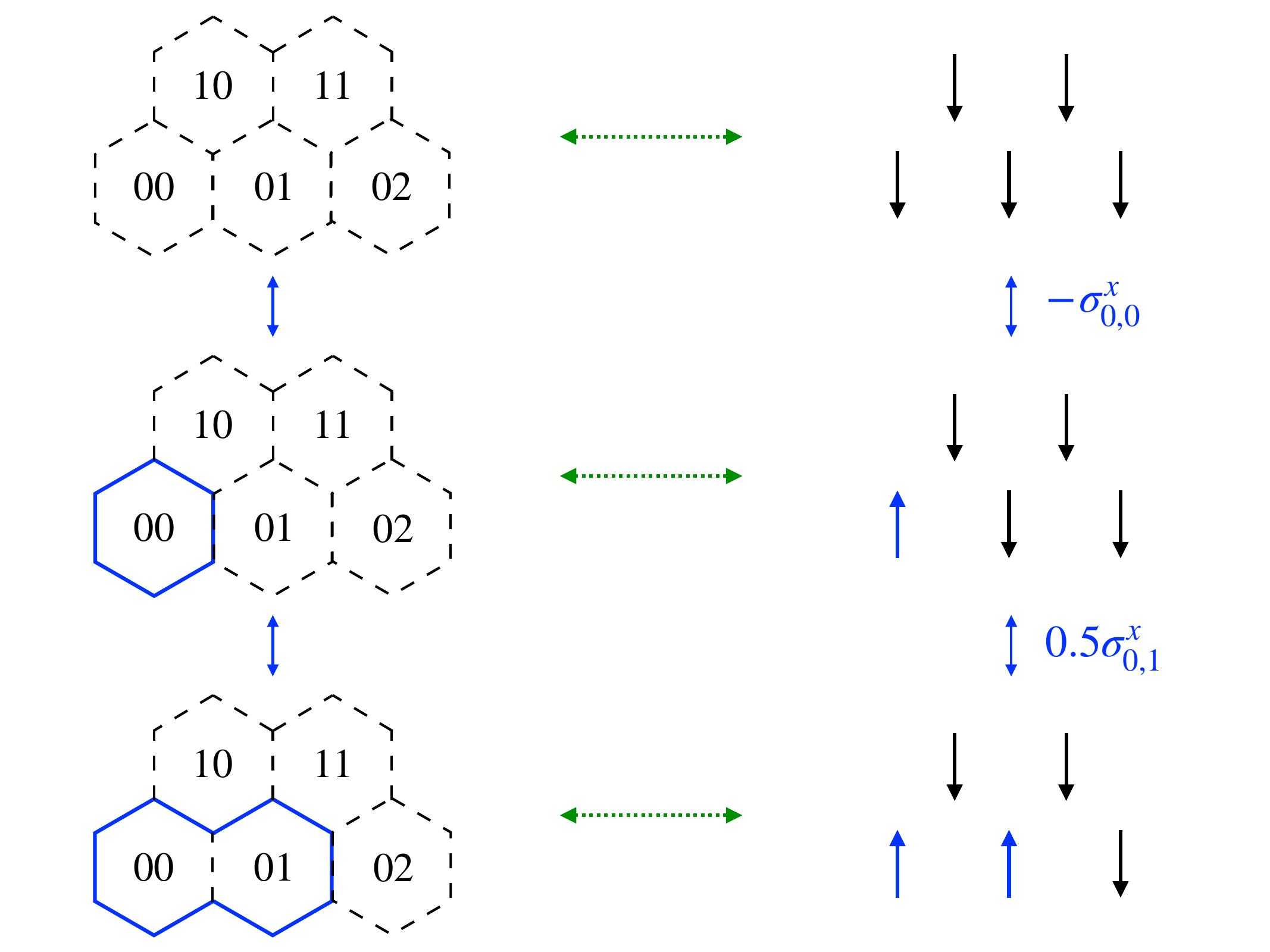}
\caption{Bijective map between physical states in $2+1$D SU(2) lattice gauge theory with a truncation of $j_{\rm max}=\frac{1}{2}$ and spin states in 2 spatial dimensions. Black dashed lines on the left represent link variables in the electric basis $j=0$ while blue solid lines stand for link variables in $j=\frac{1}{2}$. On the left, a plaquette operator at $(0,0)$ is first applied, followed by the application of another plaquette operator at $(0,1)$, which results in a two-plaquette Wilson loop with the joint link variable in $j=0$. In the spin model, these two plaquette operators correspond to two $\sigma^x$ Pauli matrices at the site $(0,0)$ and $(0,1)$ with coefficients as shown. The coefficients can be obtained from Eqs.~\eqref{eq:6j} and~\eqref{eq:vertex_1/2}.}
\label{fig:map}
\end{figure}

In the truncated Hilbert space there exist four different vertex states labeled as $0,A,B,C$ satisfying Gauss's law, which we denote by their link representations ${\bs j}\equiv(j_a,j_x,j_b)$ where $a,b$ are the internal links at the vertex and $x$ is the external link, as shown in Fig.~\ref{fig:honeycomb}: 
\begin{align}
{\bs j}_0 = (0,0,0), \quad 
{\bs j}_A = \left(\frac{1}{2},0,\frac{1}{2}\right), \quad
{\bs j}_B = \left(\frac{1}{2},\frac{1}{2},0\right), \quad
{\bs j}_C = \left(0,\frac{1}{2},\frac{1}{2}\right) \, .
\end{align}
The nonvanishing reduced matrix elements of $M_V$ between these vertex states are:
\begin{align}
\label{eq:vertex_1/2}
\langle {\bs j}_A|M_V|{\bs j}_0 \rangle 
= \langle {\bs j}_0|M_V|{\bs j}_A \rangle
= \langle {\bs j}_B|M_V|{\bs j}_C \rangle = -i, \qquad
\langle {\bs j}_C|M_V|{\bs j}_B \rangle = \frac{i}{2} \, .
\end{align}
The asymmetry of the (BC) and (CB) vertices may look strange, but if we use Eq.~\eqref{eq:vertex_1/2} to work out matrix elements of the plaquette operator on the honeycomb lattice, we find that the matrix elements are real and symmetric. This is because under the Gauss's law constraint any plaquette state has an even number $(0,2,4,6)$ of external links with $j=\frac{1}{2}$, and the numbers of (BC) and (CB) type vertices are the same.

All states that can be reached from the ground state are obtained by applying the plaquette operators $\varhexagon_{i,j}$ where $i$ and $j$ denote the plaquette position as shown in Fig.~\ref{fig:map}. For $N$ plaquettes the dimension of this truncated Hilbert space is $2^{N}$ since the repeated application of the plaquette operator to the same plaquette results in the identity: $\varhexagon_{i,j}^2=1$. Plaquette operators for different plaquettes commute: $\varhexagon_{i,j} \varhexagon_{i',j'} = \varhexagon_{i',j'} \varhexagon_{i,j}$. It is convenient to map this space onto the Hilbert space for $N$ spins where the spin-down configuration is assigned to the plaquette ground state and the spin-up configuration denotes that the plaquette operator has been applied to the plaquette \cite{Yao:2023pht}, as shown in Fig.~\ref{fig:map}. The KS Hamiltonian on the truncated Hilbert space can then be expressed in terms of the Pauli matrices $\sigma_{i,j}^\nu$, where $i,j$ denote the plaquette positions and $\nu=x,y,z$. The electric Hamiltonian is represented in terms of $\sigma_{i,j}^z$'s while the magnetic part is represented by $\sigma_{i,j}^x$'s, with coefficients determined by Eqs.~\eqref{eq:6j} and~\eqref{eq:vertex_1/2}.

A straightforward but cumbersome calculation shows that the KS Hamiltonian can be mapped onto the following Ising Hamiltonian:
\begin{align}
\label{eq:H_Ising}
aH & = h_+ \sum_{(i,j)} \Pi^+_{i,j} - h_{++} \sum_{(i,j)} \Pi^+_{i,j} \Big( \Pi^+_{i+1,j} + \Pi^+_{i,j+1} + \Pi^+_{i+1,j-1} \Big) + h_x \sum_{(i,j)} (-0.5)^{c_{i,j}}\sigma_{i,j}^x \,,
\end{align}
where $\Pi^\pm_{i,j} = (1 \pm \sigma^z_{i,j})/2$ are the projection operators onto the spin-up and spin-down states of the plaquette at $(i,j)$, respectively. The sum contains terms that refer to spins that lie outside the boundary of the system; these need to be fixed by imposing appropriate boundary conditions (see below). The coefficients in Eq.~\eqref{eq:H_Ising} are given by
\begin{align}
h_+ & = \frac{27\sqrt{3}}{8}ag^2 \,, \quad 
h_{++} = \frac{9\sqrt{3}}{8}ag^2 \,,\quad 
h_x = \frac{4\sqrt{3}}{9ag^2} \,,\nn\\
c_{i,j} & = \Pi^+_{i,j+1}\Pi^-_{i+1,j} + \Pi^+_{i+1,j}\Pi^-_{i+1,j-1} + \Pi^+_{i+1,j-1}\Pi^-_{i,j-1} \nn\\
& + \Pi^+_{i,j-1}\Pi^-_{i-1,j} 
+ \Pi^+_{i-1,j}\Pi^-_{i-1,j+1} + \Pi^+_{i-1,j+1}\Pi^-_{i,j+1} \, .
\end{align}
The expression of $c_{i,j}$ can be compactly written as
\begin{align}
\label{eq:cij}
c_{i,j} = \sum_{K=0}^5 \Pi^+_K \Pi^-_{K+1} \,,
\end{align}
where the index $K$ comes from a periodic ($K \!\mod 6$) chain $\{K=0: (i,j+1),\ K=1: (i+1,j),\ K=2: (i+1,j-1), \ K=3: (i, j-1), \ K=4: (i-1,j), \ K=5: (i-1, j+1) \}$.

We note in passing that an equivalent way of writing the magnetic part of the Hamiltonian is
\begin{align}
\label{eq:m_pauli}
h_x \sum_{(i,j)} \sigma_{i,j}^x \prod_{K=0}^5 \bigg[ \Big(\frac{1}{2}-\frac{i}{2\sqrt{2}}\Big) \sigma_K^z\sigma_{K+1}^z + \frac{1}{2}+\frac{i}{2\sqrt{2}}\bigg] \,,
\end{align}
which is more straightforward to implement on a quantum computer than the last term in Eq.~\eqref{eq:H_Ising}, since it is a local product of Pauli matrices. Expanding the product in Eq.~\eqref{eq:m_pauli} leads to multiple terms of the form $\sigma^z_K\cdots \sigma^z_{K'}\sigma_{i,j}^x$ in which there are an even number of $\sigma^z$ matrices. The generic quantum circuit to implement the time evolution driven by these terms is known~\cite{DBLP:books/daglib/0046438}: One first applies Hadamard gates $h$ to convert $\sigma^x$ rotation to $\sigma^z$ rotation $h\sigma^x h = \sigma^z$ and then realizes the generic rotation $e^{-i\theta \sigma^z_1\cdots \sigma_m^z}$ by two sequences of Controlled-NOT (CNOT) gates and a single qubit $\sigma^z$ rotation 
\begin{eqnarray}
    {\rm CNOT}(1,2){\rm CNOT}(2,3)&\cdots &{\rm CNOT}(m-1,m)e^{-i\theta\sigma_m^z}{\rm CNOT}(m-1,m) 
\nonumber\\ &\times &
    {\rm CNOT}(m-2,m-1)\cdots {\rm CNOT}(1,2),
\end{eqnarray}
where ${\rm CNOT}(i,j)$ denotes a CNOT gate on the $i$-th and $j$-th qubits with the $i$-th one as the control.

\subsubsection{Closed (Confining) Boundary Condition}

One natural choice of boundary condition is to demand that all external links of plaquettes at the boundary of the system that fall outside the boundary are in the singlet representation $j=0$. This corresponds to the requirement that there exist no gauge electric fields perpendicular to the boundary, $\vec{n}\cdot\vec{E} = 0$, which is used in the MIT-bag model to impose color confinement \cite{Johnson:1975zp}. We call these boundary conditions closed or confining. In the mapping to the 2D Ising model, the equivalent boundary condition is the requirement that all spins outside the boundary are pointing downward, i.e., their expectation values of $\sigma^z$ are $-1$. 

\subsubsection{Periodic Boundary Condition}
\label{sect:periodic}
With a periodic boundary condition, the Hamiltonian in Eq.~\eqref{eq:H_Ising} can be further simplified by shifting the energy reference point resulting in the following expression
\begin{align}
\label{eq:H_periodic}
aH = J \sum_{(i,j)} \sigma^z_{i,j}(\sigma^z_{i+1,j} + \sigma^z_{i,j+1} + \sigma^z_{i+1,j-1}) + h_x \sum_{(i,j)} (-0.5)^{c_{i,j}} \sigma^x_{i,j} \equiv JH_{zz} + h_xH_x \,,
\end{align}
with $J=-\frac{9\sqrt{3}ag^2}{32}$ and $h_x=\frac{4\sqrt{3}}{9ag^2}$. The coefficient $c_{i,j}$ is the same as in Eq.~\eqref{eq:cij}, and the magnetic part is equivalent to Eq.~\eqref{eq:m_pauli}. Unlike the case of a plaquette chain, the Hamiltonian in Eq.~\eqref{eq:H_periodic} has no linear terms in $\sigma_{i,j}^z$.

\begin{figure}
\centering
\includegraphics[width=0.7\textwidth]{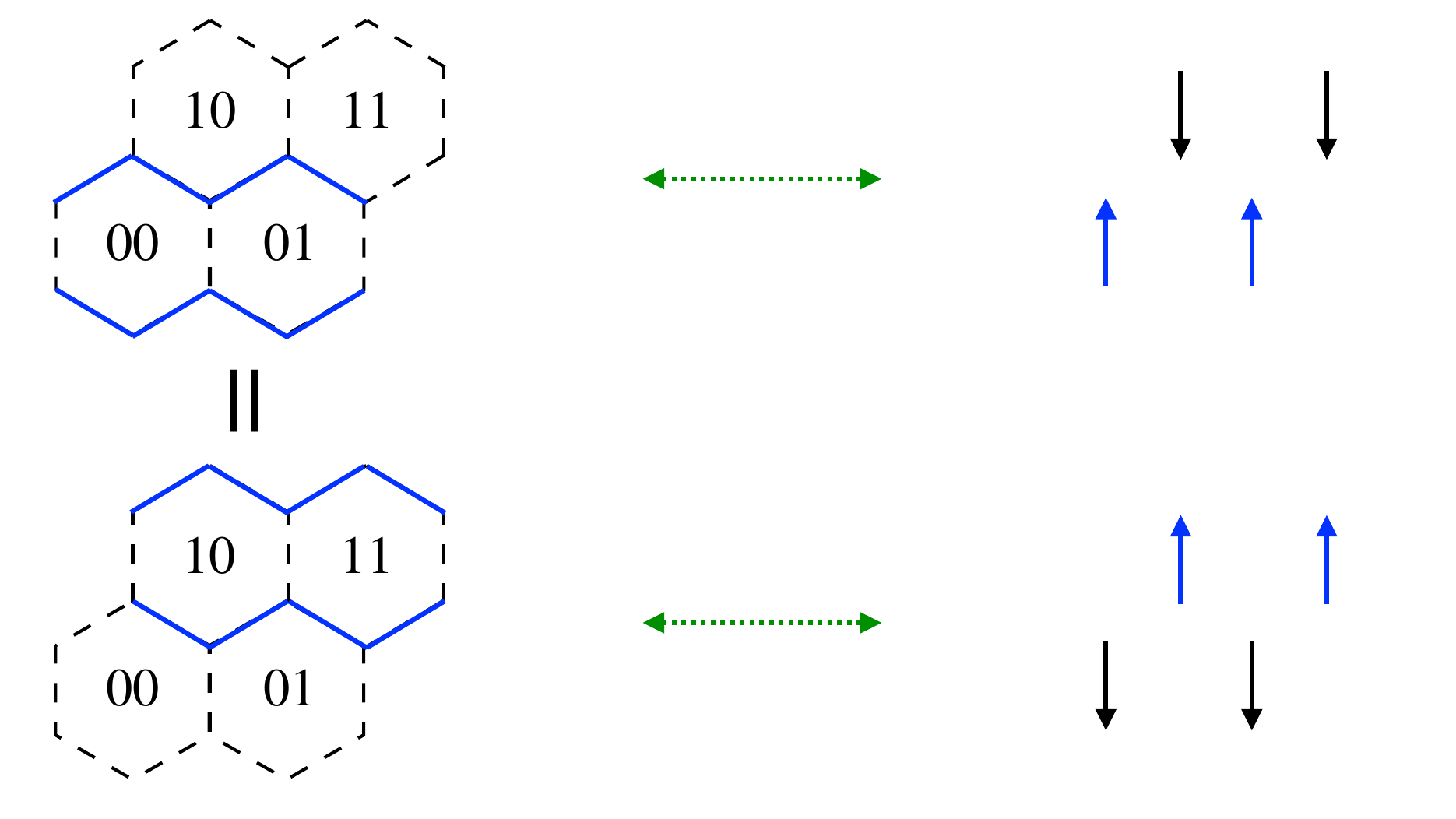}
\caption{Two equivalent states on a $2\times2$ honeycomb lattice with periodic boundary condition are mapped onto two different spin states. This redundancy in the spin representation needs to be removed.}
\label{fig:redundency}
\end{figure}

One subtlety with a periodic boundary condition is an overall spin flip redundancy in numerating states. In other words, two spin states that are related by an overall spin flip are different in the spin model, but they correspond to the same state in the original 2+1D SU(2) lattice gauge theory
\begin{align}
\prod_{(i,j)} \sigma_{i,j}^x |s\rangle \equiv |s\rangle \,.
\end{align}
This redundancy can be seen e.g., on a periodic $2\times2$ lattice from Fig.~\ref{fig:redundency}, where two different spin states are plotted and they are the same state in terms of the link variables. This redundancy needs to be removed in practical calculations.

\subsubsection{Representation of Wilson Loops}

Often in practical applications, one wants to evaluate expectation values of gauge invariant operators such as Wilson loops. Here we list the representation of 1-plaquette and 2-plaquette Wilson loops in the spin model discussed above.

For a 1-plaquette Wilson loop operator, the spin model representation is the same as one term in the magnetic part of the Hamiltonian. If the 1-plaquette operator $O_1$ is located at $(i,j)$, its expression is
\begin{align}
O_1 \equiv - \sigma_{i,j}^x \prod_{K=0}^5 \bigg[ \Big(\frac{1}{2}-\frac{i}{2\sqrt{2}}\Big) \sigma_K^z\sigma_{K+1}^z + \frac{1}{2} + \frac{i}{2\sqrt{2}}\bigg] \,,
\end{align}
where the index $K$ comes from a periodic ($K \!\mod 6$) chain $\{K=0: (i,j+1),\ K=1: (i+1,j),\ K=2: (i+1,j-1), \ K=3: (i, j-1), \ K=4: (i-1,j), \ K=5: (i-1, j+1) \}$.

For a 2-plaquette Wilson loop operator $O_2$ located at $(i,j),(i,j+1)$, its representation in the spin model is given by
\begin{align}
O_2 &\equiv - \sigma_{i,j}^x \sigma_{i,j+1}^x \frac{1+3\sigma_{i,j}^z\sigma_{i,j+1}^z}{4} \prod_{K=0}^7 (-0.5)^{\Pi^+_K \Pi^-_{K+1}} 
\nn\\
& = - \sigma_{i,j}^x \sigma_{i,j+1}^x \frac{1+3\sigma_{i,j}^z\sigma_{i,j+1}^z}{4} \prod_{K=0}^7 \bigg[ \Big(\frac{1}{2}-\frac{i}{2\sqrt{2}}\Big) \sigma_K^z\sigma_{K+1}^z + \frac{1}{2} + \frac{i}{2\sqrt{2}}\bigg] \,,
\end{align}
where the index $K$ here forms a periodic ($K \!\mod 8$) chain $\{K=0: (i,j+2),\ K=1: (i+1,j+1),\ K=2: (i+1,j), \ K=3: (i+1, j-1), \ K=4: (i, j-1), \ K=5: (i-1, j), \ K=6: (i-1, j+1)\ K=7: (i-1,j+2),  \}$.

\section{Momentum States and Matrix Elements with Periodic Boundary Condition}
\label{sec:k-sector}

With the periodic boundary condition, the spin model on the honeycomb lattice has translational invariance in three directions orthogonal to the three unit vectors $\hat{e}_i$. Only two of them are linearly independent. We choose them to be $x$ and $y$ directions: $\hat{x}\equiv(1,0)$ and $\hat{y}\equiv (\frac{1}{2}, \frac{\sqrt{3}}{2})$. Then we have
\begin{align}
[\hat{T}_x, H] = 0\,, \qquad [\hat{T}_y, H] = 0 \,,
\end{align}
where $\hat{T}_i$ denotes the translation operator along the $i$-th spatial direction by one lattice unit. The vanishing commutators mean we can simultaneously diagonalize the Hamiltonian and translation operators. The eigenstates of the two translation operators are states with specific momenta, which can be written as
\begin{align}
\label{eqn:kstate}
|a(k_x,k_y)\rangle = \frac{1}{\sqrt{N_a}} \sum_{r_x=0}^{N_x-1} \sum_{r_y=0}^{N_y-1} e^{-ik_xr_x-ik_yr_y} \hat{T}_x^{r_x} \hat{T}_y^{r_y} |a\rangle \,,
\end{align}
where $|a\rangle$ is a representative state in the translational equivalent classes defined by $\hat{T}_x$ and $\hat{T}_y$. We assume there are $N_x$ plaquettes along the $x$ direction and $N_y$ along the $y$ direction. The momenta are given by $k_x = 2\pi n_x/N_x, k_y = 2\pi n_y/N_y$ where $n_x\in\{1,2,\cdots,N_x-1\}$ and $n_y\in\{1,2,\cdots,N_y-1\}$. The normalization factor $N_a$ can be numerically obtained by sweeping through all the states involved in Eq.~\eqref{eqn:kstate} and calculating the norm of the linear superposition state accounting for the redundancy discussed in Section~\ref{sect:periodic}. If the two sums on the right hand side of Eq.~\eqref{eqn:kstate} vanish, then the corresponding momentum state does not exist. The state $|a(k_x,k_y)\rangle$, whenever exists, is an eigenstate of $\hat{T}_i$ with eigenvalue $e^{ik_i}$ for $i=x,y$.

The matrix element of the electric part of Eq.~\eqref{eq:H_periodic} in the momentum state basis is given by
\begin{align}
\langle b(k'_x,k'_y) | H_{zz} | a(k_x,k_y) \rangle = & ~ \delta_{ab}\delta_{k_xk_x'}\delta_{k_yk_y'}\sum_{i=0}^{N_x-1}\sum_{j=0}^{N_y-1} \nn\\
& \Big( z_{i,j}(a) z_{i,j+1}(a) + z_{i,j}(a) z_{i+1,j}(a) + z_{i,j}(a) z_{i+1,j-1}(a) \Big) \,,
\end{align}
where $z_{i,j}(a) = \pm 1$ is the eigenvalue of the operator $\sigma_{i,j}^z$ applied to the state $|a\rangle$.

The matrix element of the magnetic part of Eq.~\eqref{eq:H_periodic} in the momentum state basis can be worked out similarly
\begin{align}
&  \langle b(k'_x,k'_y) | H_x | a(k_x,k_y)\rangle =  \nn\\
& \ \delta_{k_xk_x'}\delta_{k_yk_y'} \sqrt{\frac{N_b}{N_a}} \sum_{i=0}^{N_x-1}\sum_{j=0}^{N_y-1} e^{-ik_x\ell_i - ik_y\ell_j} \prod_{K=0}^5 \bigg[ \Big(\frac{1}{2}-\frac{i}{2\sqrt{2}}\Big) z_K(a) z_{K+1}(a) + \frac{1}{2} + \frac{i}{2\sqrt{2}}\bigg]  \,,
\end{align}
where the index $K$ depends on $(i,j)$ and is given as above. The integers $\ell_i$ and $\ell_j$ are determined by
\begin{align}
\sigma_{i,j}^x |a\rangle = \hat{T}_x^{-\ell_i} \hat{T}_y^{-\ell_j} |b\rangle \,.
\end{align}

Finally, we give explicit expressions for the matrix elements of the 1-plaquette and 2-plaquette operators $O_1$ and $O_2$ in the momentum eigenstate basis. Without loss of generality, we assume the $O_1$ operator acts at $(i,j)=(0,0)$ and the $O_2$ operator acts on the plaquette pair at $(0,0),(0,1)$. Similarly as for the matrix element of the magnetic Hamiltonian, we find
\begin{align}
\langle b(k_x',k_y')| O_1 | a(k_x,k_y) \rangle =& - \frac{1}{N_xN_y}\sqrt{\frac{N_b}{N_a}}  \sum_{r_x=0}^{N_x-1} \sum_{r_y=0}^{N_y-1} e^{i\phi(\vec{k},\vec{k}';r_x,r_y)} \\
&\times \prod_{K=0}^5 \bigg[ \Big(\frac{1}{2}-\frac{i}{2\sqrt{2}}\Big) z_K(a) z_{K+1}(a) + \frac{1}{2} + \frac{i}{2\sqrt{2}}\bigg] \,, \nn
\end{align}
where the indices $K$ denote the six plaquettes surrounding $(-r_x,-r_y)$, the phase $\phi$ is 
\begin{align}
\phi(\vec{k},\vec{k}';r_x,r_y) = (k_x'-k_x)r_x+(k_y'-k_y)r_y-k_x'\ell_{r_x} -k_y'\ell_{r_y} \,,
\end{align}
and the two integers $\ell_{r_x}$ and $\ell_{r_y}$ are determined by the condition
\begin{align}
\sigma_{-r_x,-r_y}^x |a\rangle = \hat{T}_x^{-\ell_{r_x}} \hat{T}_y^{-\ell_{r_y}} |b\rangle \,.
\end{align}
A similar construction gives the matrix element of $O_2$ as
\begin{align}
\langle b(k_x',k_y')| O_2 | a(k_x,k_y) \rangle = & - \frac{1}{N_xN_y}\sqrt{\frac{N_b}{N_a}} \sum_{r_x=0}^{N_x-1} \sum_{r_y=0}^{N_y-1} e^{i\phi(\vec{k},\vec{k}';r_x,r_y)} \frac{1+3z_{-r_x,-r_y}(a) z_{-r_x,1-r_y}(a) }{4} \nn \\
& \times \prod_{K=0}^7 \bigg[ \Big(\frac{1}{2}-\frac{i}{2\sqrt{2}}\Big) z_K(a) z_{K+1}(a) + \frac{1}{2} + \frac{i}{2\sqrt{2}}\bigg]   \,,
\end{align}
where the indices $K$ denote the eight plaquettes around $(-r_x,-r_y),(-r_x,1-r_y)$, and $\ell_{r_x}$ and $\ell_{r_y}$ are determined by
\begin{align}
\sigma_{-r_x,-r_y}^x \sigma_{-r_x,1-r_y}^x |a\rangle = \hat{T}_x^{-\ell_{r_x}} \hat{T}_y^{-\ell_{r_y}} |b\rangle \,.
\end{align}

The momentum basis and the corresponding matrix elements may not be very useful for quantum simulation, but they have been applied in studies of testing ETH for non-Abelian gauge theories~\cite{Yao:2023pht} and turned out to be very useful to enlarge the system size accessible on a classical computer.

\section{Conclusions}
\label{sec:conclusions}
In this work we construct a simple 2D spin Hamiltonian that exactly describes the $2+1$ dimensional SU(2) lattice theory with a truncation at $j_{\rm max}=\frac{1}{2}$ in the electric basis. During the construction, Gauss's law is fully accounted for, and the Hilbert space of the spin model only contains physical states of the original SU(2) theory. Although Gauss's law is fully implemented, the interactions in the spin model are still local. As a result, the system can be efficiently simulated on quantum hardware that does not have all-to-all connectivity. Therefore, the simple 2D spin Hamiltonian can serve as a benchmark for future quantum simulation studies of lattice gauge theory that tests their robustness against unphysical state admixture caused by hardware noise and algorithm errors. Furthermore, this simple spin Hamiltonian allows us to exactly diagonalize it for reasonably large system sizes, which is necessary to test the ETH for a subset of the physical Hilbert space of the original SU(2) theory. We will pursue this in future work.

\acknowledgments
BM thanks Natalie Klco for useful discussions. XY is supported by the U.S. Department of Energy, Office of Science, Office of Nuclear Physics, InQubator for Quantum Simulation (IQuS) (https://iqus.uw.edu) under Award Number DOE (NP) Award DE-SC0020970 via the program on Quantum Horizons: QIS Research and Innovation for Nuclear Science. BM is supported by grant DE-FG02-05ER41367 from the U.S. Department of Energy.

\bibliography{main}
\end{document}